\pdfoutput=1
\RequirePackage{fix-cm}
\documentclass[sigconf,screen=true]{acmart}

\usepackage[utf8]{inputenc}
\usepackage[T1]{fontenc}

\usepackage{subfig}
\usepackage{natbib}
\usepackage{balance}

\usepackage{fancyhdr}

\newcommand{\hk}[1]{{#1}} 
\newcommand{\ts}[1]{{#1}} 
\newcommand{\trs}[1]{{#1}} 
\newcommand{\rmv}[1]{{#1}} 

\renewcommand\footnotetextcopyrightpermission[1]{} 

\copyrightyear{2019}
\acmYear{2019}
\setcopyright{acmlicensed}
\acmConference[IVA '19]{ACM International Conference on Intelligent Virtual Agents}{July 2--5, 2019}{Paris, France}
\acmBooktitle{ACM International Conference on Intelligent Virtual Agents (IVA '19), July 2--5, 2019, Paris, France}
\acmPrice{15.00}
\acmDOI{10.1145/3308532.3329472}
\acmISBN{978-1-4503-6672-4/19/07}

\hypersetup{pdftitle={Analyzing Input and Output Representations for Speech-Driven Gesture Generation},%
 pdfauthor={Taras Kucherenko, Dai Hasegawa, Gustav Eje Henter, Naoshi Kaneko, Hedvig Kjellström},%
 pdfkeywords={Gesture generation, social robotics, representation learning, neural network, deep learning, virtual agents}} 

\begin{document}

\title[Analyzing Input and Output Representations for Speech-Driven Gesture Generation]{Analyzing Input and Output Representations\\for Speech-Driven Gesture Generation}

%
\author{Taras Kucherenko}
\affiliation{\institution{KTH Royal Institute of Technology\\
Stockholm, Sweden}}
\email{tarask@kth.se}
\author{Dai Hasegawa}
\affiliation{\institution{Hokkai Gakuen University\\ Sapporo, Japan}}
\email{dhasegawa@hgu.jp} 

\author{Gustav Eje Henter}
\affiliation{\institution{KTH Royal Institute of Technology\\
Stockholm, Sweden}}
\email{ghe@kth.se}
\author{Naoshi Kaneko}
\affiliation{\institution{Aoyama Gakuin University\\Sagamihara, Japan}}
\email{kaneko@it.aoyama.ac.jp}
%

\author{Hedvig Kjellstr{\"o}m}
\affiliation{\institution{KTH Royal Institute of Technology\\
Stockholm, Sweden}}
\email{hedvig@kth.se}

%
\renewcommand{\shortauthors}{T.\ Kucherenko \emph{et al.}}

%
\begin{abstract}

This paper presents a novel framework for automatic speech-driven
gesture generation, applicable to human-agent interaction including both virtual agents and robots. Specifically, we extend recent deep-learning-based,
data-driven methods for speech-driven gesture generation by
incorporating representation learning.
Our model takes speech as input and produces gestures as output, in the form of a sequence of 3D coordinates. 

Our approach consists of two steps. First, we learn a lower-dimensional
representation of human motion using a denoising autoencoder
neural network, consisting of a motion encoder \textit{MotionE} and a motion decoder \textit{MotionD}. The learned representation preserves the most important aspects of the human pose variation while removing less relevant variation. Second, we train a novel encoder network \textit{SpeechE} to map from speech
to a corresponding motion representation with reduced dimensionality. At test time, the speech encoder and the motion decoder networks are combined: \textit{SpeechE} predicts motion representations based on a given speech signal and \textit{MotionD} then decodes these representations to produce motion sequences.

We evaluate different representation sizes in order to find the most effective dimensionality for the representation. We also evaluate the effects of using different speech features as input to the model. We find that mel-frequency cepstral coefficients (MFCCs), alone or combined with prosodic features, perform the best. The results of a subsequent user study confirm the benefits of the representation learning.

\end{abstract}

%
%

%

%

%
\maketitle 


\vspace{-1mm}
\section{Introduction}
\label{sec:intro}
Conversational agents in the form of virtual agents or social robots are rapidly becoming wide-spread \rmv{ and many of us will soon interact regularly with them in our day-to-day lives}. Humans use non-verbal behaviors to signal their intent, emotions and attitudes in human-human interactions \cite{knapp2013nonverbal,matsumoto2013nonverbal}. Similarly, it has been shown that people read and interpret robots' non-verbal cues similarly to non-verbal cues from other people \cite{breazeal2005effects}. Robots that are equipped with such non-verbal behaviors have shown to positively affect people's perception of the robot \cite{salem2013err}.
Conversational agents therefore need the ability to perceive and produce non-verbal communication.

An important part of non-verbal communication is \ts{gesticulation: gestures made with hands, arms, head pose and body pose communicate a large share of non-verbal content} \cite{mcneill1992hand}. \hk{To facilitate natural human-agent interaction, i}t is hence important to enable robots \hk{and embodied virtual agents} to accompany their speech with gestures in the way people do.

Most existing work on generating hand gestures relies on rule-based methods \cite{cassell2001beat, ng2010synchronized, huang2012robot}. \ts{These methods are rather rigid as they
can only generate gestures that are incorporated in the rules. Writing down rules for all possible gestures found in human interaction is highly labor-intense and time-consuming. Consequently, it is difficult to fully capture the richness of human gesticulation in rule-based systems.}
In this paper, we present a solution that eliminates this \hk{bottleneck} by using a data-driven method that learns to generate human gestures from a dataset of human actions. 
More specifically, we use speech data, as it is highly correlated with hand gestures \cite{mcneill1992hand} and has the same temporal character.


\ts{To predict gestures from speech, we apply Deep Neural Networks (DNNs), which have been widely used in human skeleton modeling for motion prediction \cite{martinez2017human} as well as classification \cite{butepage2017deep}. We further apply representation learning on top of conventional speech-input, gesture-output DNNs.}
Representation learning is a branch of unsupervised learning 
aiming to learn a better representation of the data.
Typically, representation learning is applied to make a subsequent learning task easier.
Inspired by previous successful applications to 
learning human motion dynamics, for example in prediction \cite{butepage2017deep} 
and motion synthesis \cite{habibie2017recurrent}, this paper applies representation learning to the motion sequence, in order to extend previous approaches for neural-network-based speech-to-gesture mappings \cite{takeuchi2017speech,hasegawa2018evaluation}.

The contributions of this paper are \hk{two}-fold: 
\vspace{-1mm}
\begin{enumerate}
    \item We propose a novel speech-driven non-verbal behavior generation \hk{method} that \trs{can be applied to any embodiment}.
    \item We evaluate the importance of representation both for the motion (by doing representation learning) and for the speech (by comparing different speech feature extractors).
\end{enumerate}
\vspace{-1mm}
We analyze which motion representation size yields the best results for the speech-driven gesture generation.  Moreover, we numerically evaluate which speech features are most useful. Finally, we perform a user study, which finds that representation learning improved the perceived naturalness of the gestures over the baseline model.

\trs{Our work here extends our previous publication \cite{kucherenko2019importance} by expanding the method description, investigating additional input features and significantly widening the scope of the objective evaluation.} A video summary of this paper with visual examples is available at \href{https://youtu.be/Iv7UBe92zrw}{youtu.be/Iv7UBe92zrw}.


\section{Representation learning for speech-motion mapping}
\label{sec:method}

\begin{figure}
\includegraphics[width=0.8\linewidth]{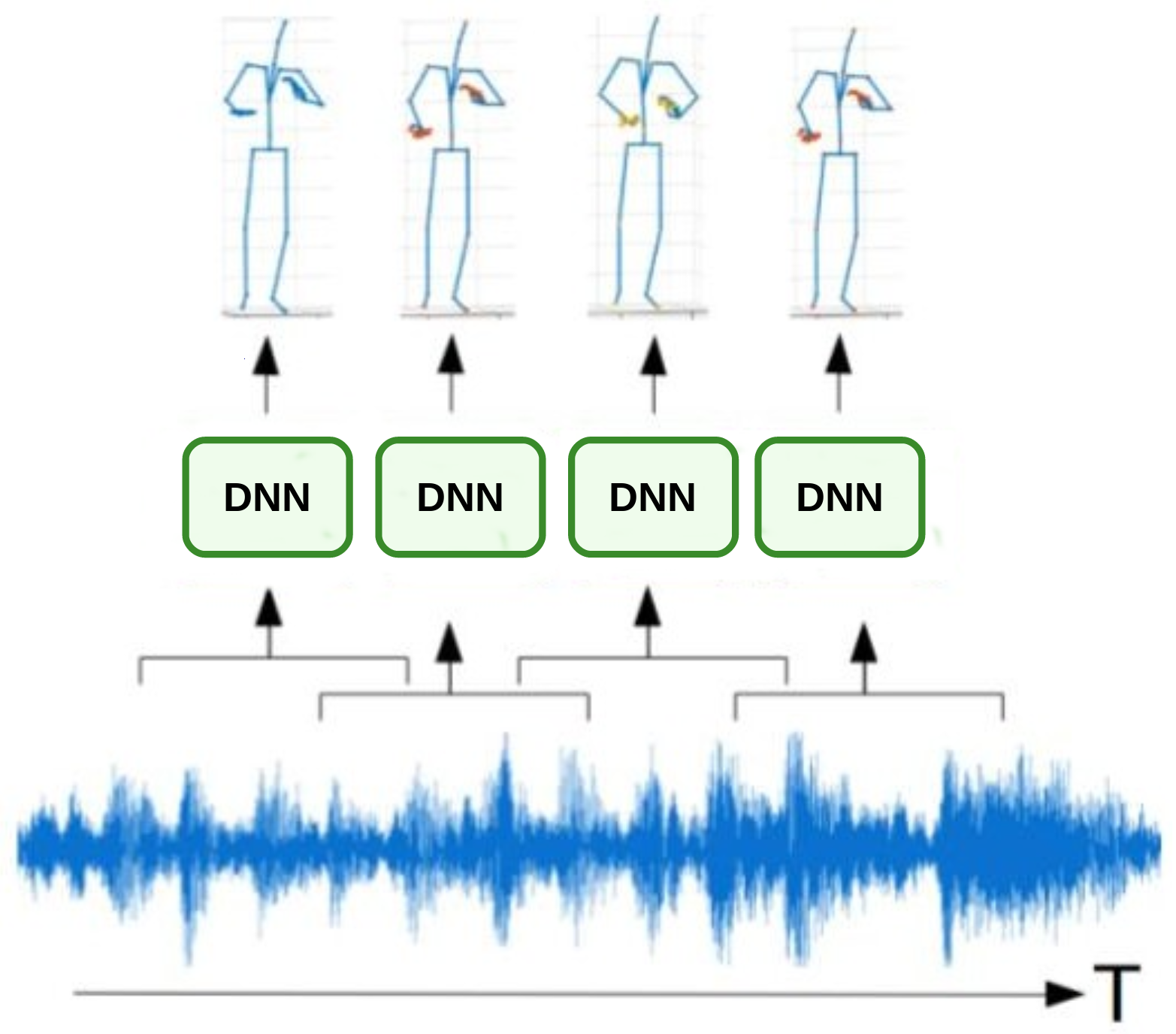}
\vspace{-2mm}
\caption{Framework overview. \hk{The Deep Neural Network (DNN) green boxes are further described in Figures \ref{fig:baseline_model} and \ref{fig:our_model}}.}
\label{fig:framework_overview}
\vspace{-3mm}
\end{figure}

\subsection{Problem formulation}

We frame the problem of speech-driven gesture generation as follows: given a sequence of speech features $\boldsymbol{s} = [\boldsymbol{s}_t]_{t=1:T}$ extracted from segments (frames) of speech audio at regular intervals $t$, the task is to generate a corresponding gesture sequence $\boldsymbol{\hat{g}} = [\boldsymbol{\hat{g}}_t]_{t=1:T}$ that a human might perform while uttering this speech.

A speech segment $s_t$ would be typically represented by some features, such as Mel-Frequency Cepstral Coefficients \trs{\cite{davis1980comparison}}, MFCCs, (which are commonly used in speech recognition) or prosodic features including pitch (F0), energy, and their derivatives (which are commonly used in speech emotion analysis).
Similarly, the ground truth gestures $g_t$ and predicted gestures $\hat{g}_t$ are typically represented as 3D-coordinate sequences:
$
\boldsymbol{g}_t = [x_{i,t}, y_{i,t}, z_{i,t}] _{i=1:n}\text{,}
$
$n$ being the number of keypoints of the human body (such as shoulder, elbow, etc.) that are being modelled.

The most recent systems tend to perform mappings from $\boldsymbol{s}$ to $\hat{\boldsymbol{g}}$ using a neural network (NN) learned from data. The dataset typically contains recordings of human motion (for instance from a motion capture system) and the corresponding speech signals.

\subsection{Baseline speech-to-motion mapping}
\label{ssec:baseline}
Our model \hk{builds} on the work of Hasegawa et al.~\cite{hasegawa2018evaluation}. In this section, we describe their model, which is our baseline system.

The speech-gesture neural network \cite{hasegawa2018evaluation} takes a speech sequence as input and generates a sequence of gestures frame by frame. As illustrated in Figure \ref{fig:framework_overview}, the speech is processed in overlapping chunks of $C=30$ frames (like in \cite{hasegawa2018evaluation}) before and after the current time $t$. (The offset between frames \hk{in the figure} is exaggerated for demonstration purposes.) 
\hk{An entire} speech-feature \hk{window} is fed into the network \hk{at each time step $t$}: $\mathrm{NN}_{\mathrm{input}} = [s_{t-C}, ... s_{t-1}, s_t, s_{t+1}, ... s_{t+C}]$.
The network is regularized by predicting not only the \hk{pose} but also the velocity as output: $\mathrm{NN}_{\mathrm{output}} = [g_{t},\Delta g_{t}]$. While incorporating the velocity into test-time predictions did not provide a significant improvement, the inclusion of velocity as a multitask objective during training forced the network to learn motion dynamics \cite{takeuchi2017speech}.

\begin{figure}
\includegraphics[width=0.98\linewidth]{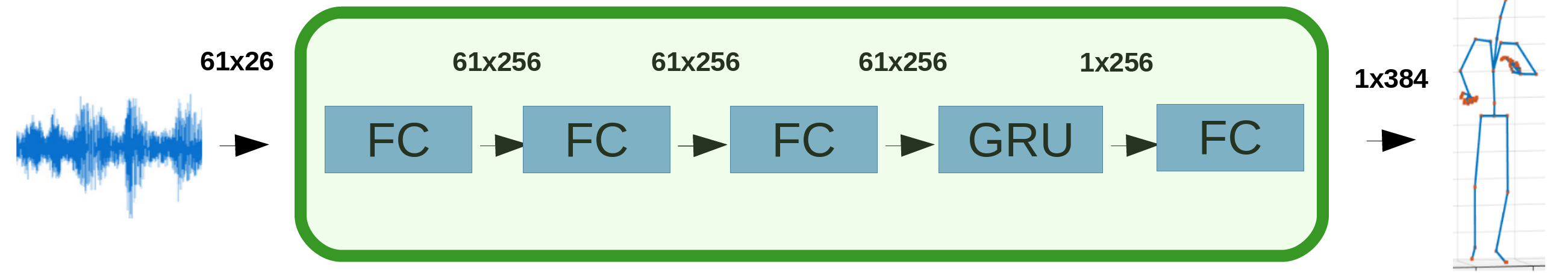}
\caption{Baseline DNN for speech-to-motion mapping. The green box identifies the part used for the DNN in Figure \ref{fig:framework_overview}.}
\label{fig:baseline_model}
\vspace{-5mm}
\end{figure} 

The \textbf{baseline neural network architecture} is illustrated in Figure \ref{fig:baseline_model}. First, MFCC features are computed for every speech segment. Then three fully connected layers (FC) are applied to every chunk $s_t$. 
Next, a recurrent network layer with Gated Recurrent Units (GRUs) \cite{cho2014properties} is applied to the resulting sequence. 
Finally, an additional linear, fully-connected layer is used as the output layer.

\hk{We note that} the baseline network we described is a minor modification of the network in \cite{hasegawa2018evaluation}. Specifically, we use a different type of recurrent network units, namely GRUs instead of B-LSTMs. \ts{Our experiments found that} this cuts the training time in half while maintaining the same prediction performance. We also used shorter window length for computing MFCC features, namely 0.02 s instead of 0.125 s, since MFCCs were developed to be informative about speech for these window lengths. The only other difference against \cite{hasegawa2018evaluation} is that we did not post-process (smooth) the output sequences.


\subsection{Proposed approach}
\label{ssec:proposed}

Our intent is to extend the baseline model by leveraging the power of representation learning.
Our proposed approach has three steps:
\begin{enumerate}
    \item We apply representation learning to learn a motion representation \hk{$\boldsymbol{z}$}.
    \item We learn a mapping from the chosen speech features \hk{$\boldsymbol{s}$} to the learned motion representation \hk{$\boldsymbol{z}$} (using the same NN architecture as in the baseline model).
    \item The two learned mappings are chained together to turn speech input \hk{$\boldsymbol{s}$} into motion output \hk{$\boldsymbol{g}$}.
\end{enumerate}

\begin{figure}[t]
\centering
\vspace{-2mm}
\subfloat[MotionED: Representation learning for the motion\label{sfig:motioned}]{\hfill\includegraphics[width=0.8\linewidth]{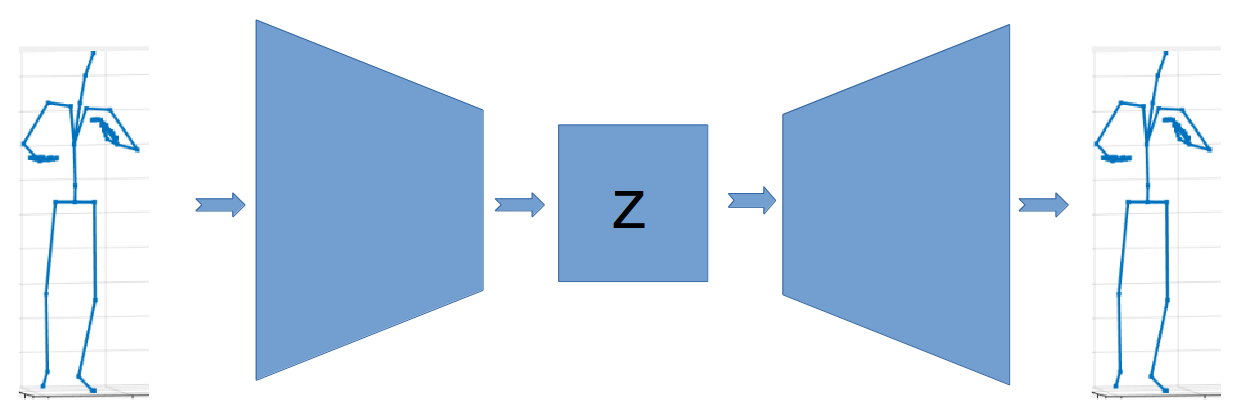}\hfill}\\
\subfloat[SpeechE: Mapping speech to motion representations\label{sfig:speeche}]{\hfill\includegraphics[width=0.8\linewidth]{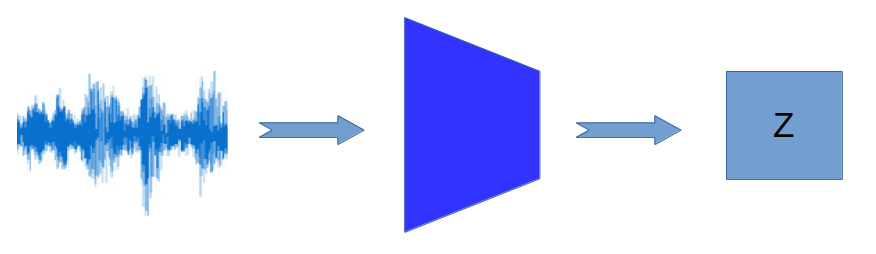}\hfill}\\
\subfloat[Combining the learned components: SpeechE and MotionD\label{sfig:combined}]{\hfill\includegraphics[width=0.8\linewidth]{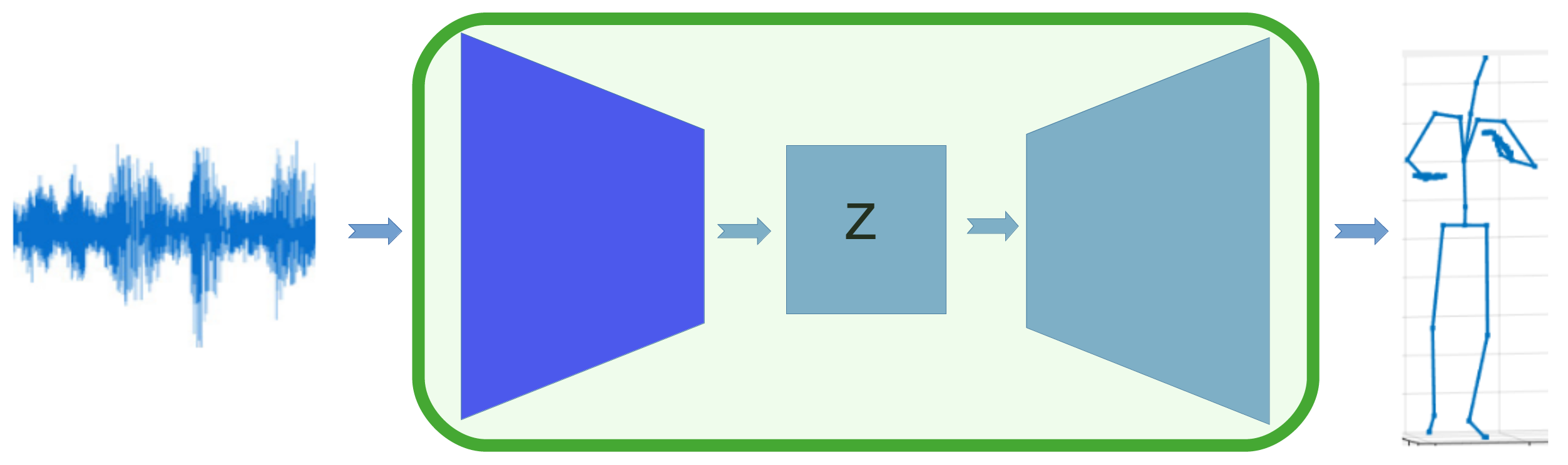}\hfill}
\vspace{-2mm}
\caption{How the proposed encoder-decoder DNN for speech-to-motion mapping is constructed. The green box \hk{denotes} the part of the system used for the DNN in Figure \ref{fig:framework_overview}.}
\label{fig:our_model}
\vspace{-4mm}
\end{figure}

\noindent\textbf{Motion representation learning}

\noindent Figure \ref{sfig:motioned} illustrates representation learning for human motion sequences. The aim of this step is to reduce the motion dimensionality, which confers two benefits: 1) simplifying the learning problem by reducing the output space dimensionality; and 2) reducing redundancy in the training data by forcing the system to concentrate important information to fewer numbers.

To learn motion representations, we used a neural network structure called a Denoising Autoencoder (DAE) \cite{vincent2010stacked} with one hidden layer ($\boldsymbol{z}$).
This network learns to reconstruct the original data from input examples with additive noise while having a bottleneck layer in the middle. This bottleneck forces the network to compute lower dimensional representation.
The network can be seen as a combination of two networks: \textit{MotionE}, which encodes the motion $\boldsymbol{m}$ to the representation $\boldsymbol{z}$ and \textit{MotionD}, which decodes the representation $\boldsymbol{z}$ back to the motion $\boldsymbol{m}$:
\begin{align}
\boldsymbol{z} & = MotionE(\boldsymbol{m})\\
\hat{\boldsymbol{m}} & = MotionD(\boldsymbol{z})
\end{align}
The neural network learns to reconstruct the original motion coordinates as closely as possible by minimizing the mean squared error (MSE) loss function: $\mathrm{MSE}(\boldsymbol{m},\hat{\boldsymbol{m}}) = \lVert \hat{\boldsymbol{m}} - \boldsymbol{m}\rVert_2^2$.

\medskip
\noindent\textbf{\hk{Encoding speech to the motion representation}}

\noindent Figure \ref{sfig:speeche} illustrates the principle of how we map speech to motion representation. Conceptually, the network performing this task fulfills the same role as the baseline network in Section \ref{ssec:baseline}. 
The main difference versus the baseline is that the output of the network is not raw motion values, but a compact, learned representation of motion.
To be as comparable as possible to the baseline, we use the same network architecture to map speech to motion representations in the proposed system as the baseline used for mapping speech to motion. We call this network \textit{SpeechE}.

\medskip
\noindent\textbf{\hk{Connecting the speech encoder and the motion decoder}}

\noindent Figure \ref{sfig:combined} illustrates how the system is used at testing time by chaining together the two previously learned mappings. First, speech input is fed to the \textit{SpeechE} \hk{encoding net}, which produces a sequence of motion representations. Those motion representations are then decoded into joint coordinates by the \textit{MotionD} \hk{decoding net}.

\break

\subsection{Implementation}
\textbf{The baseline neural network}

\noindent Figure \ref{fig:baseline_model} shows the structure and layer sizes of the neural network used in the baseline system. As seen, the network inputs contained 61 $\times$ 26 elements, comprising 26-dimensional speech-derived MFCC vectors from the current frame plus 30 adjacent frames both before and after it, resulting in a total of 61 vectors in the input (While we describe and explore other audio features in \ref{sec:technical_details}, the baseline model only used MFCCs, to be consistent with \cite{hasegawa2018evaluation}).
The Fully Connected (FC) layers and the Gated Recurrent Unit (GRU) layers both had a width of 256 and used the ReLU activation function. Batch normalization and dropout with probability 0.1 of dropping activations were applied between every layer. Training minimized the mean squared error between predicted and ground-truth gesture sequences using the Adam optimizer \cite{kingma2014adam} with learning rate 0.001 and batch size 2048. Training was run for 120 epochs, after which no further improvement in validation set loss was observed. \trs{Save for batch size and the number of epochs these hyperparameters were taken from the baseline paper \cite{hasegawa2018evaluation}.}

\medskip
\noindent\textbf{The denoising autoencoder neural network}


\noindent We trained a DAE with input size 384 (64 joints: 192 3D-coordinates and their first derivatives) and one hidden, feedforward layer in the encoder and decoder. 
Hyperparameters were optimized on our validation dataset, described in Section \ref{ssec:dataset}.
Different widths were investigated for the bottleneck layer (see Section \ref{ssec:dim_analysis}), with 325 units giving the best valiadation-data performance. Gaussian noise was added to each input with a standard deviation equal to 0.05 times the standard deviation of that feature dimension. Training minimized the MSE reconstruction loss using Adam with a learning rate of 0.001 and batch size 128. Training was run for 20 epochs.



\section{Experimental setup}
\label{sec:exp_setup}

This section describes the data and gives technical detail regarding the experiments we conducted to evaluate the importance of input and output representations in speech-driven gesture generation.

\subsection{Gesture-speech dataset}
\label{ssec:dataset}
For our experiments, we used a gesture-speech dataset collected by Takeuchi et al.\ \cite{takeuchi2017creating}. \ts{Motion data were recorded in a motion capture studio from two Japanese individuals having a conversation in the form of an interview.} An experimenter asked questions prepared beforehand, and a performer answered them. The dataset contains MP3-encoded speech audio captured using headset microphones on each speaker, coupled with motion-capture motion data stored in the BioVision Hierarchy format (BVH). \rmv{The BVH data describes motion as a time sequence of Euler rotations for each joint in the defined skeleton hierarchy.} 
These Euler angles were converted to a total of 64 global joint positions in 3D. As some recordings had a different framerate than others, we downsampled all recordings to a common framerate of to 20 frames per second (fps).
\trs{For the representation learning,} each dimension was standardized to mean zero and maximum (absolute) value one.


The dataset contains 1,047 utterances\footnote{The original paper reports 1,049 utterances, which is a typo.}, of which our experiments used 957 for training, 45
for validation, and 45 testing.  
The relationship between various speech-audio features and the 64 joint positions was thus learned from 171 minutes of training data at 20 fps, resulting in 206,000 training frames.

\subsection{Feature extraction}
\label{sec:technical_details}

The ease of learning and the limits of \ts{expressiveness} for a speech-to-gesture system \ts{depend} greatly on the input features used. Simple features that encapsulate the most important information are likely to work well for learning from small datasets, whereas rich and complex features might allow learning additional aspects of speech-driven gesture behavior, but may require more data to achieve good accuracy. We experimented with three different, well-established audio features as inputs to the neural network, namely i) MFCCs, ii) spectrograms, and iii) prosodic features.


In terms of implementation, 26 MFCCs were extracted with a window length of 0.02 s and a hop length of 0.01 s, which amounts to 100 analysis frames per second. Our spectrogram features, meanwhile, were 64-dimensional and extracted with the window length and hop size 0.005 s, yielding a rate of 200 fps. Frequencies \ts{that carry} little speech information (below the hearing threshold of 20 Hz, or above 8000 Hz) were removed. Both the MFCC and the spectrogram sequences were downsampled to match the motion frequency of 20 fps by replacing every 5 (MFCCs) or 10 (spectrogram) frames by their average. (This averaging prevents aliasing artifacts.)

As an alternative to MFCCs and spectrum-based features, we also considered prosodic features. These differ in that prosody encompasses intonation, rhythm, and anything else about the speech outside of the specific words spoken (e.g., semantics and syntax). Prosodic features were previously used for gesture prediction in early data-driven work by Chiu \& Marsella \cite{chiu2011train}. For this study, we considered pitch and energy (intensity) information. The information in these features has a lower bitrate and is not sufficient for discriminating between and responding differently to arbitrary words, but may still be informative for predicting non-verbal emphases like beat gestures and their timings.

We considered four specific prosodic features, extracted from the speech audio with a window length of 0.005 s, resulting in 200 fps. Our two first prosodic features were the energy of the speech signal and the time derivative (finite difference) of the energy series. The third and fourth features were the logarithm of the F0 (pitch) contour, which contains information about the speech intonation, and its time derivative. We extracted pitch and intensity values from audio using Praat \cite{boersma2002praat} and normalized pitch and intensity as in \cite{chiu2011train}: the pitch values were adjusted by taking $\log(x+1)-4$ and setting negative values to zero, and the intensity values were adjusted by taking $\log(x)-3$. All these features were again downsampled to the motion frequency of 20 fps using averaging. 



\subsection{Numerical evaluation measures}
\label{ssec:evaluation_measures}

We used both objective and subjective measures to evaluate the different approaches under investigation.
Among the former, two kinds of error measures were considered:
\begin{description}
\item [Average Position Error (APE)] The APE is the average Euclidean distance between the predicted coordinates $\hat{g}$ and the original coordinates $g$:
\begin{equation}
\label{eq:task}
\mathrm{APE}(g_t^n,\hat{g}_t^n) = \frac{1}{DT} \sum_{t=1}^T \sum_{d=1}^D \lVert g_t^n - \hat{g}_t^n \rVert_2
\end{equation}
where $T$ is the total duration of the sequence, $D$ is the dimensionality of the motion data and \textit{n} is a sequence index. 
\item [Motion Statistics]
We considered the average values and distributions of acceleration and jerk for the produced motion.
\end{description}

We believe the motion statistics to be the most informative for our task: in contrast to tracking, the purpose of gesture generation is not to reproduce one specific \textit{true} position, but rather to produce a plausible candidate for natural motion. Plausible motions do not require measures like speed or jerk to closely follow the original motion, but they should follow a similar distribution. That is why we study distribution statistics, namely average speed and jerk.

Since there is some randomness in system training, e.g., due to random initial network weights, we evaluated every condition five times and report the mean and standard deviation of those results.

\section{Results and Discussion}
\label{sec:exper_res}
This section presents an analysis of the performance of the gesture-prediction system.
We investigate different design aspects of our system that relate to the importance of representations, namely the speech and motion representations used.

\subsection{Importance of motion encoding}
\label{ssec:dim_analysis}
\begin{figure}[t]
\centering
\subfloat[Average position error (APE). The baseline APE (blue line) is 8.3$\mathbf{\pm}$0.4.\label{sfig:ape}]{\includegraphics[width=0.99\linewidth]{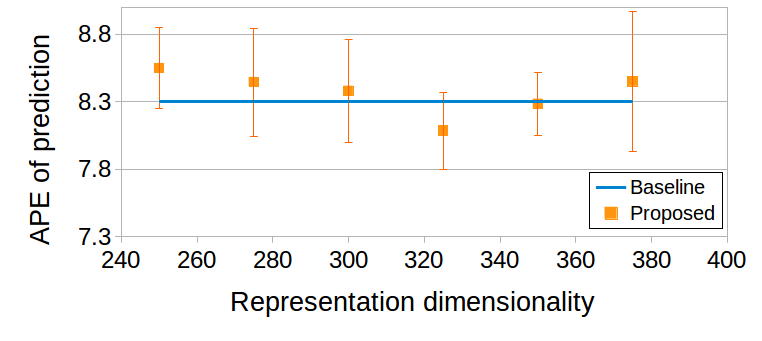}}\\
\subfloat[Average jerk. Baseline jerk is 2.8$\mathbf{\pm}$0.3 while ground-truth jerk is 0.54.\label{sfig:avg_jerk}]{\includegraphics[width=0.99\linewidth]{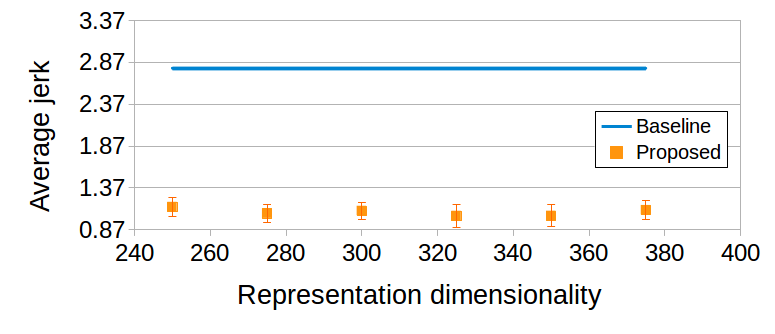}}\\
\vspace{-2mm}
\caption{Effect of learned-representation dimensionality in the proposed model.}
\label{fig:dim_analysis}
\end{figure}

We first evaluated how different dimensionalities for the learned motion representation affected the prediction accuracy of the full system.
Figure \ref{fig:dim_analysis} graphs the results of this evaluation. In terms of average position error (APE) (see Figure \ref{sfig:ape}) the optimal embedding-space dimensionality is clearly 325, which is smaller than the original data dimensionality (384). Motion jerkiness (see Figure \ref{sfig:avg_jerk}) is also lowest for dimensionality 325, but only by a slight margin compared to the uncertainty in the estimates. Importantly, the proposed system performs much better than the baseline \cite{hasegawa2018evaluation} on both evaluation measures. The difference in the average jerk, in particular, is highly significant. This validates our decision to use representation learning to improve gesture generation models. While motion jerkiness can be reduced through post-processing, as in \cite{hasegawa2018evaluation}, that does not address the underlying shortcoming of the model.

The numerical results are seen to vary noticeably between different runs, suggesting that training might converge to different local optima depending on the random initial weights.

\subsection{Input speech representation}

Having established the benefits of representation learning for the output motion, we next analyze which input features perform the best for our speech-driven gesture generation system. In particular, we compare three different features -- MFCCs, raw power-spectrogram values, and prosodic features (log F0 contour, energy, and their derivatives) -- as described in Section \ref{sec:technical_details}.

From Table \ref{tab:speech_features}, we observe that MFCCs achieve the lowest APE, but produce motion with higher acceleration and jerkiness than the spectrogram features do. Spectrogram features gave suboptimal APE, but match ground-truth acceleration and jerk better than the other features we studied.


\begin{table}
  \caption{\hk{Objective evaluation of different speech features, averaged over five re-trainings of the system.}}
  \label{tab:speech_features}
  \begin{tabular}{lccc}
    \toprule
    Model/feature & APE & Acceleration & Jerk \\
    \toprule
    Static mean pose & 8.95 & 0 & 0\\
    \midrule
    Prosodic & 8.56$\pm${0.2}  & 0.90$\pm${0.03} & 1.52$\pm${0.07}\\
    Spectrogram  &  8.27$\pm${0.4}  & \textbf{0.51}$\pm${0.07} & \textbf{0.85}$\pm${0.12}\\
    Spectr.\ + Pros. & 8.11$\pm${0.3}  & 0.57$\pm${0.08} & 0.95$\pm${0.12}\\
    MFCC & \textbf{7.66}$\pm${0.2} & 0.53$\pm${0.03} & 0.91$\pm${0.05} \\
    MFCC + Pros. & \textbf{7.65}$\pm${0.2} & 0.58$\pm${0.06} & 0.97$\pm${0.11}\\
    \midrule
    Baseline \cite{hasegawa2018evaluation} (MFCC) & 8.07$\pm${0.1}  & 1.50$\pm${0.03}   & 2.62$\pm${0.05}\\
    \midrule
    Ground truth & 0 & 0.38 & 0.54
\end{tabular}
\end{table}

\subsection{Detailed performance analysis}
The objective measures in Table \ref{tab:speech_features} do not unambiguously establish which input features would be the best choice for our predictor. We therefore further analyze the statistics of the generated motion, particularly acceleration. Producing the right motion with the right acceleration distribution is crucial for generating convincing motions, as too fast or too slow motion does not look natural.

\begin{figure}[t]
\centering
\vspace{-2mm}
\subfloat[Average acceleration histogram.\label{sfig:avg_acc}]{\hfill\includegraphics[width=0.9\linewidth]{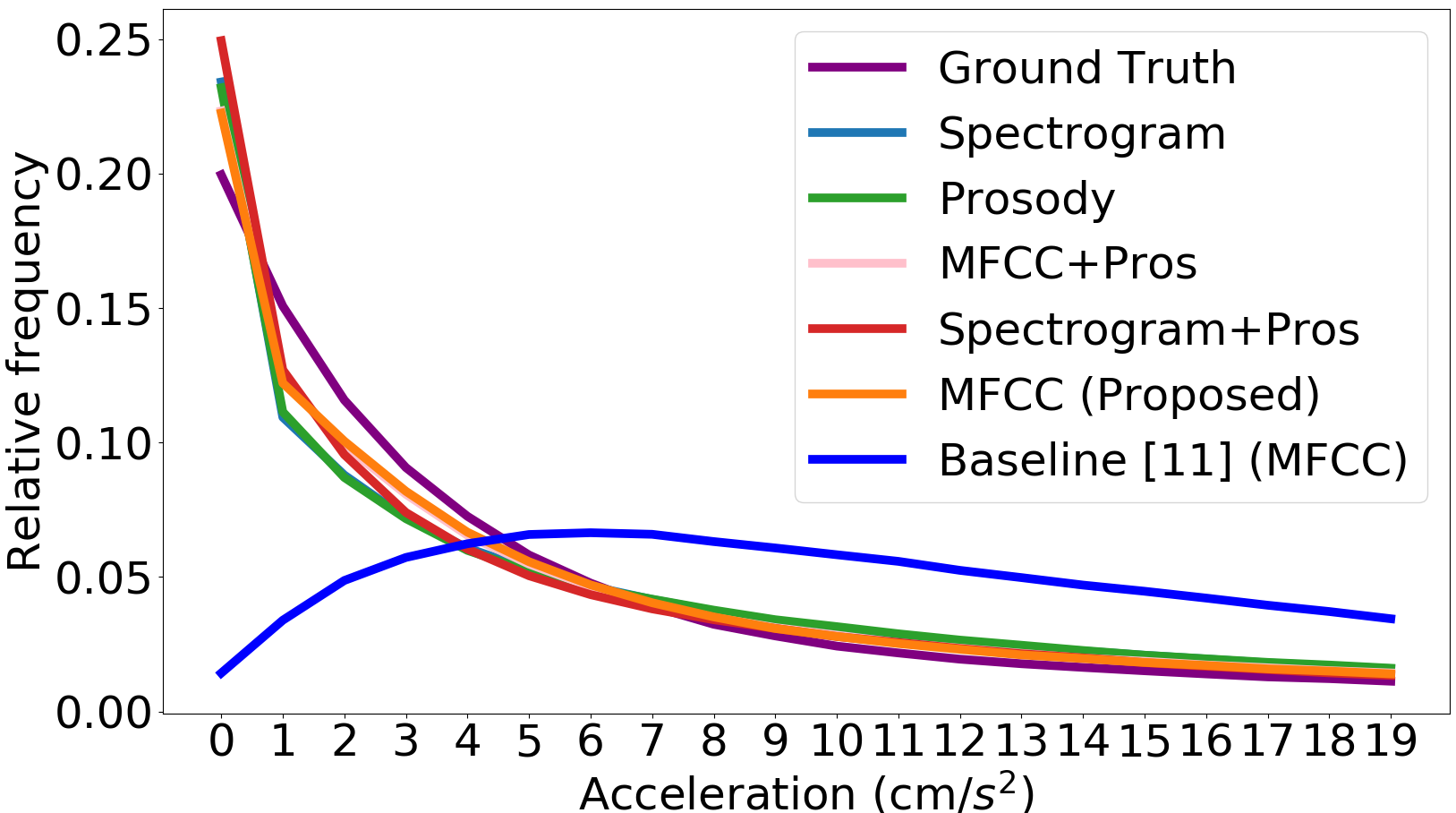}\hfill}\\
\subfloat[Acceleration histogram for shoulders. Legend as in (a).\label{sfig:shoulders_acc}]{\hfill\includegraphics[width=0.9\linewidth]{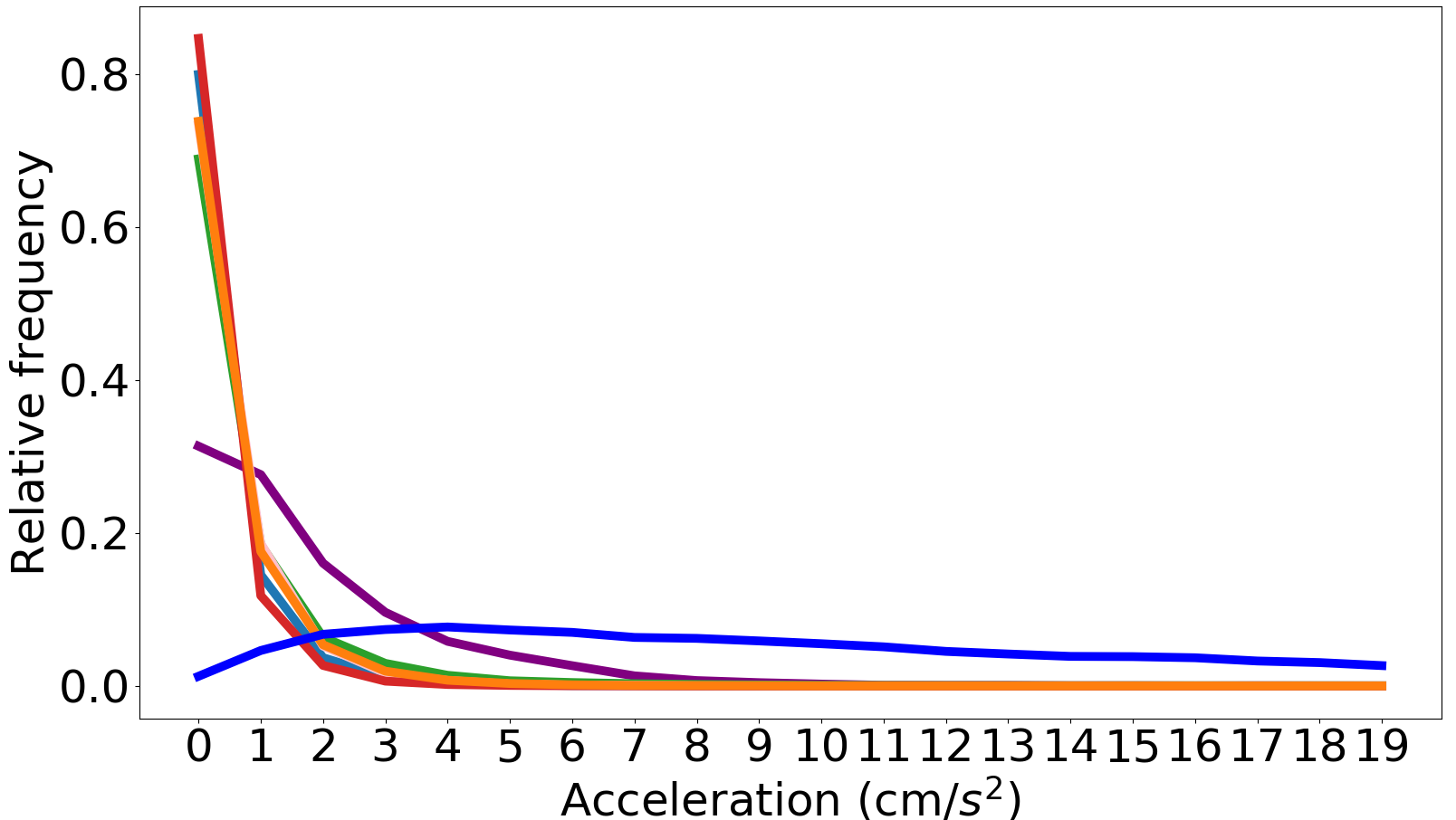}\hfill}\\
\subfloat[Acceleration histogram for hands Legend as in (a).\label{sfig:hands_acc}]{\hfill\includegraphics[width=0.9\linewidth]{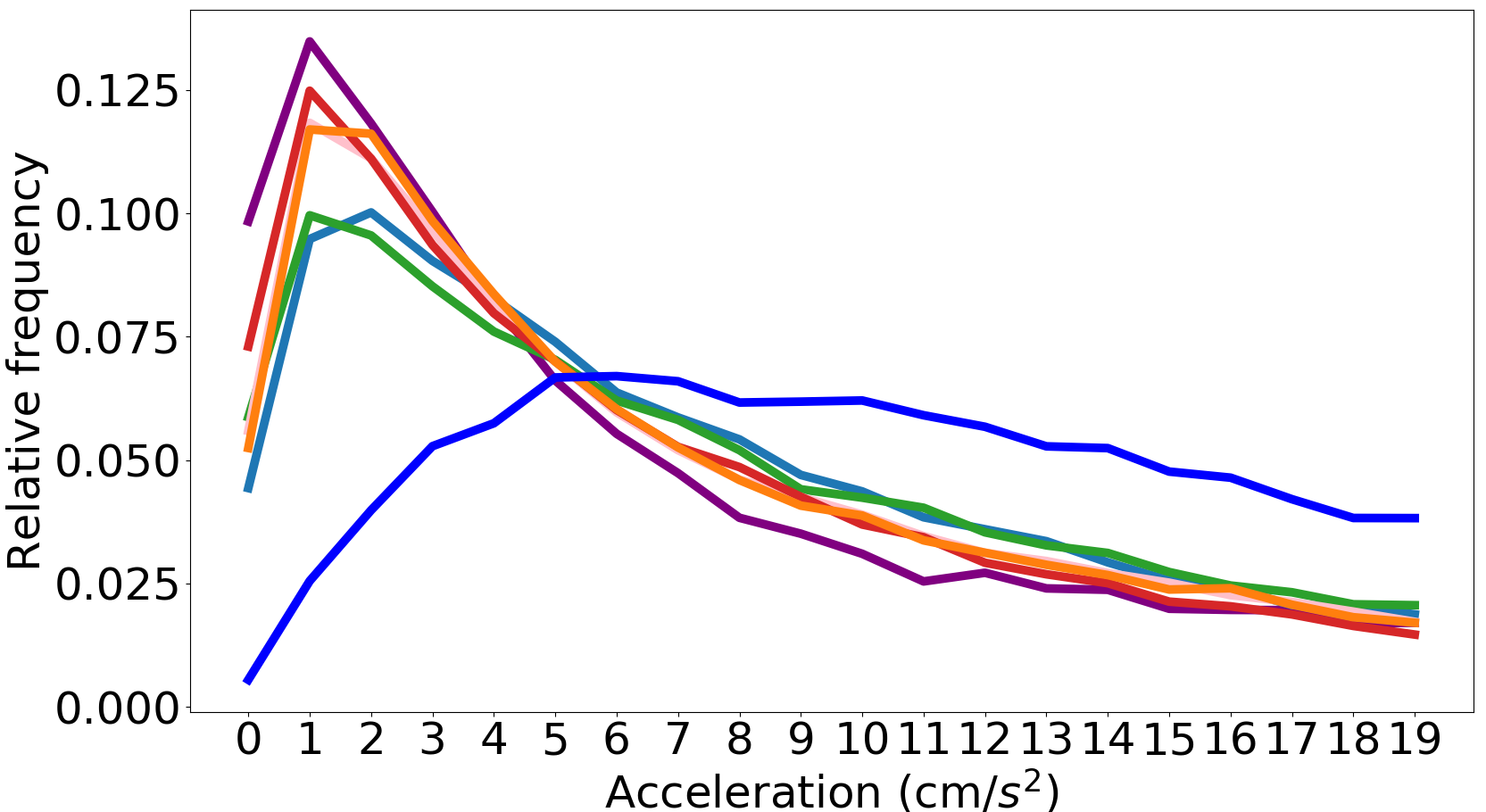}\hfill}\\
\caption{Acceleration distributions given different speech features. 
Firstly, the motion produced from our model (with any input feature) is more similar to the acceleration distribution of the ground-truth motion, compared to motion from the baseline model. Secondly, we find that MFCCs produce an acceleration distribution most similar to the ground truth, especially for the hands, as shown in (c).}
\label{fig:speed_hist}
\end{figure}

\ts{To investigate the motion statistics associated with the different input features, we computed acceleration histograms of the generated motions and compared those against histograms derived from the ground truth}. We calculated the relative frequency of different acceleration values over frames in all 45 test sequences, split into bins of equal width. 
\ts{For easy comparison, our histograms are visualized as line plots rather than bar plots.}  

Figure \ref{sfig:avg_acc} presents acceleration histograms across all joints for different input features. The acceleration distribution of the baseline model 
deviate more from the ground truth than our model does.

Since the results in Figure \ref{sfig:avg_acc} are averaged over all joints, they do not indicate whether 
all the joints move naturally. To address this we also analyze the acceleration distribution for certain specific joints.
Figure \ref{sfig:shoulders_acc} shows an acceleration histogram calculated for the shoulders only. We see that our system with all the speech features has acceleration distributions very close to one another, but that all of them far away from the actual data.
A possible explanation for this could be that shoulder motion might be difficult to predict from the speech input, in which case the predicted motion is likely to stay close the mean shoulder position. 

Figure \ref{sfig:hands_acc} shows acceleration histograms for the hands. 
Hands convey the most important gesture information, suggesting that this plot is the most informative.
Here, the MFCC-based system is much closer to the ground truth.
Combining MFCCs and prosodic features resulted in similar performance as for MFCC inputs alone. This could be due to redundancy in the information exposed by MFCCs and prosodic features, or due to our networks and optimizer not being able to exploit synergies between the two representations.

Taken together, Figures \ref{sfig:avg_acc}-c suggest that motion generated from MFCC features give acceleration statistics as similar or more similar to the ground truth as those of motion generated from other features. Moreover, using MFCCs 
as input features makes our proposed system consistent with the baseline paper \cite{hasegawa2018evaluation}.



 

\subsection{User study}
\label{ssec:user_study}


The most important goal in gesture generation is to produce motion patterns that are convincing to human observers.
Since improvements in objective measures do not always translate into superior subjective quality for human observers, we validated our conclusions by means of a user study comparing key systems.

\begin{table}
  \caption{\hk{Statements evaluated in user study}}
  \label{tab:eval_stat}
  \begin{tabular}{ll}
    \toprule
    Scale & Statement (translated from Japanese) \\
    \midrule
    Naturalness & Gesture was natural \\
      & Gesture was smooth \\
      & Gesture was comfortable \\
    \midrule
    Time &Gesture timing was matched to speech\\
     consistency& Gesture speed was matched to speech\\
     & Gesture pace was matched to speech\\
    \midrule
    Semantic&Gesture was matched to speech content\\
    consistency  & Gesture well described speech content\\
      & Gesture helped me understand the content\\
  \bottomrule
\end{tabular}
\end{table}

We conducted a 1$\times$2 factorial design with the within-subjects factor being representation learning (baseline vs.\ encoded). The encoded gestures were generated by the proposed method from MFCC input. We randomly selected 10 utterances from a test dataset of 45 utterances, for each of which we created two videos using the two gesture generation systems. Visual examples are provided at \url{https://vimeo.com/album/5667276}. After watching each video, we asked participants to rate nine statements about the naturalness, time consistency, and semantic consistency of the motion. The statements were the same as in the baseline paper \cite{hasegawa2018evaluation} and are listed in Table \ref{tab:eval_stat}. 
Ratings used a seven-point Likert scale anchored from strongly disagree (1) to strongly agree (7). The utterance order was fixed for every participant, but the gesture conditions (baseline vs.\ encoded) were counter-balanced. 
With 10 speech segments and two gesture-generation systems, we obtained 20 videos, producing 180 ratings in total per subject, 60 for each scale in Table \ref{tab:eval_stat}.

\begin{figure}[t]
\centering
\includegraphics[width=0.9\linewidth]{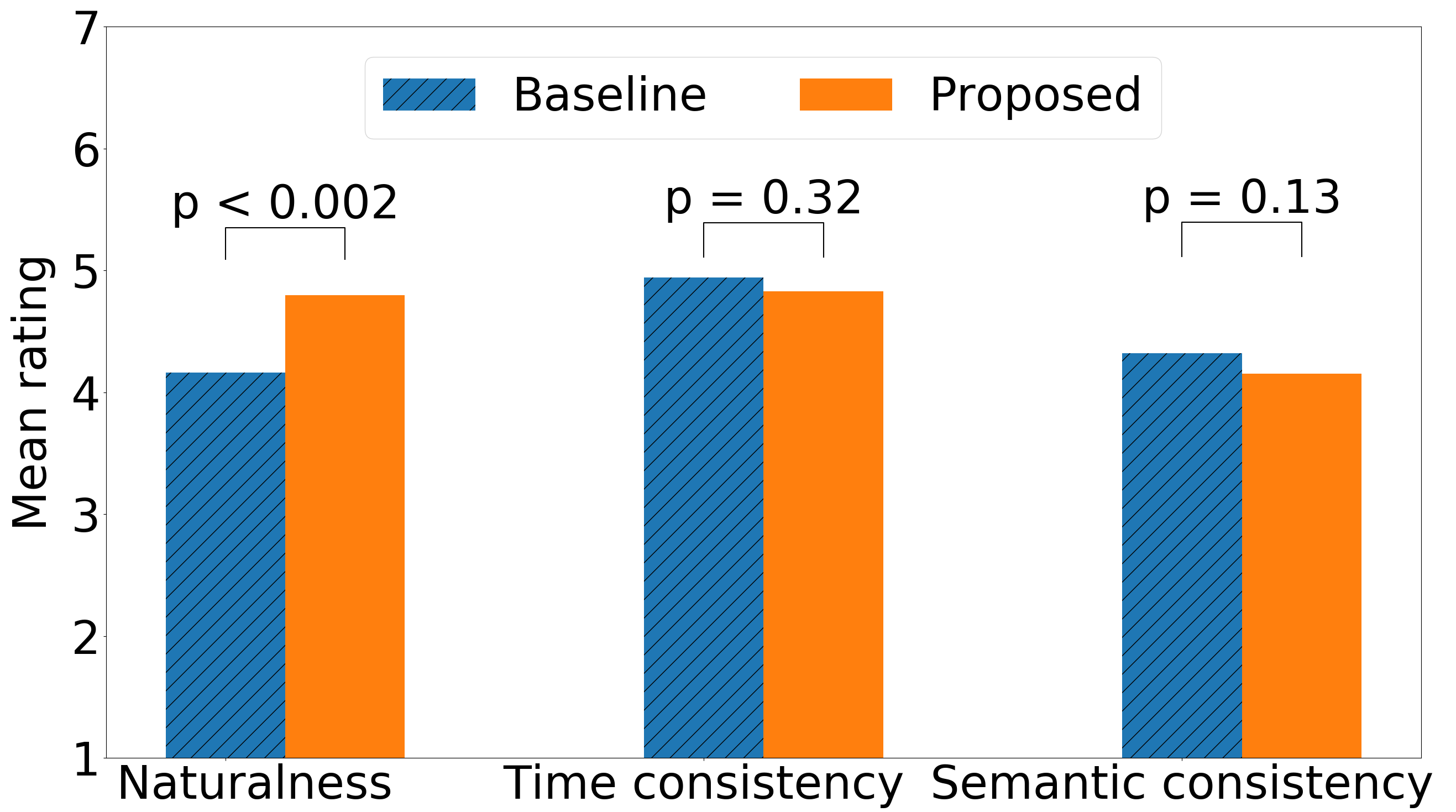}
\vspace{-1mm}
\caption{\hk{Results from the u}ser study. We note a significant difference in naturalness, but not the other scales.}
\label{fig:user_study}
\end{figure}

19 native speakers of Japanese (17 male, 2 female), on average 26 years old, participated in the user study. A paired-sample $t$-test was conducted to evaluate the impact of the motion encoding on the perception of the produced gestures. Figure \ref{fig:user_study} illustrates the results we obtained for the three scales being evaluated. We found a significant difference in naturalness between the baseline (M=4.16, SD=0.93) and proposed model (M=4.79, SD=0.89), $t$=-3.6372, $p$<0.002. A 95\%-confidence interval for the mean rating improvement with the proposed system is (0.27,1.00). There were no significant difference on the other scales: for time-consistency $t$=1.0192, $p$=0.32, for semantic consistency $t$=1.5667, $p$=0.13. These results indicate that gestures generated by the proposed method (i.e., with representation learning) were perceived as more natural than the baseline.

\section{Related work}
\label{sec:rel_work}

While most of the work on non-verbal behavior generation in the literature considers rule-based systems \cite{cassell2001beat,huang2012robot,salvi2009synface}, we review only data-driven approaches and pay special attention to methods incorporating elements of representation learning, since that is the direction of our research. For a review of rule-based systems, we refer the reader to Wagner et al.\ \cite{wagner2014gesture}.

\subsection{Data-driven head and face movements}
\label{ssec:dd_facial}
Facial-expression generation has been an active field of research for several decades. Many of the state-of-the-art methods are data-driven. Several recent works have applied neural networks in this domain \cite{haag2016bidirectional,greenwood2017predicting,sadoughi2018novel,sadoughi2017joint,suwajanakorn2017synthesizing}. Among the cited works, Haag \& Shimodaira \cite{haag2016bidirectional} use a bottleneck network to learn compact representations, although their bottleneck features subsequently are used to define prediction inputs rather than prediction outputs as in the work we presented. Our proposed method works on a different aspect of non-verbal behavior that co-occurs with speech, namely generating body motion driven by speech.  
 
\subsection{Data-driven body motion generation}
\label{ssec:dd_gestures}
Generating body motion 
is an active area of research with applications to animation, computer games, and other simulations. 
Current state-of-the-art approaches in such body-motion generation are generally data-driven and based on deep learning \cite{zhang2018mode,li2017auto,pavllo2018quaternet}. 
Zhou et al.\ \cite{li2017auto} proposed a modified training regime to make recurrent neural networks generate human motion with greater long-term stability, while Pavllo et al.\ \cite{pavllo2018quaternet} formulated separate short-term and long-term recurrent motion predictors, using quaternions to more adequately express body rotations. 

Some particularly relevant works for our purposes are \cite{liu2014feature,holden2015learning,holden2016deep,butepage2017deep}. All of these leverage representation learning (various forms of autoencoders) that predict human motion, yielding accurate yet parsimonious predictors. Habibie et al.\ \cite{habibie2017recurrent} extended this general approach to include an external control signal in an application to human locomotion generation with body speed and direction as the control input. Our approach is broadly similar, but generates body motion from speech rather than position information.

\subsection{Speech-driven gesture generation}
\label{ssec:sd_gestures}
Like body motion in general, gesture generation has also begun to shift towards data-driven methods, for example \cite{chiu2011train,chiu2015predicting,hasegawa2018evaluation}. Several researchers have tried to combine data-driven approaches with rule-based systems. For example, Bergmann \& Kopp \cite{bergmann2009GNetIc} learned a Bayesian decision network for generating iconic gestures. Their system is a hybrid between data-driven and rule-based models because they learn rules from data. Sadoughi et al.~\cite{sadoughi2017speech} used probabilistic graphical models with an additional hidden node to provide contextual information, such as a discourse function.
They experimented on only three hand gestures and two head motions. We believe that regression methods that learn and predict arbitrary movements, like the one we have proposed, represent a more flexible and scalable approach than the use of discrete and pre-defined gestures.

The work of Chiu \& Marsella \cite{chiu2011train} is of great relevance to the work have presented, in that they took a regression approach and also utilized representation learning. Specifically, they used wrist height in upper-body motion to identify gesticulation in motion capture data of persons engaged in conversation. A network based on Restricted Boltzmann Machines (RBMs) was used to learn representations of arm gesture motion, and these representations were subsequently predicted based on prosodic speech-feature inputs using another network also based on RBMs.
\trs{Levine et al. \cite{levine2010gesture} also used an intermediate state between speech and gestures. The main differences are that they used hidden Markov models, whose discrete states are less powerful than recurrent neural networks, and that they selected motions from a fixed library, while our model can generate unseen gestures.} Later Chiu et al.~\cite{chiu2015predicting} proposed a method to predict co-verbal gestures using a machine learning setup with feedforward neural networks followed by Conditional Random Fields (CRFs) for temporal smoothing. They limited themselves to a set of 12 discrete, pre-defined gestures and used a classification-based approach.

Recently, Hasegawa et al.\ \cite{hasegawa2018evaluation} designed a speech-driven neural network capable of producing 3D motion sequences. We built our model on this work while extending it with motion-representation learning, since learned representations have improved motion prediction in other applications as surveyed in Section \ref{ssec:dd_gestures}.

\section{Conclusions and future work}
\label{sec:concl}

This paper presented a new model for speech-driven gesture generation.
Our method extends prior work on deep learning for gesture generation by applying representation learning. 
The motion representation is learned first, after which a network is trained to predict such representations from speech, instead of directly mapping speech to raw joint coordinates as in prior work. We also evaluated the effect of different representations for the input speech. \trs{Our code is publicly available to encourage replication of our results.\footnote{ \href{https://github.com/GestureGeneration/Speech_driven_gesture_generation_with_autoencoder}{github.com/GestureGeneration/Speech\_driven\_gesture\_generation\_with\_autoencoder}}}

Our experiments show that representation learning improves the objective and subjective performance of the speech-to-gesture neural network. 
Although models with and without representation learning were rated similarly in terms of time consistency and semantic consistency, subjects rated the gestures generated by the proposed method as significantly more natural than the baseline.

The main limitation of our method, as with any data-driven method and particularly those based on deep learning, is \ts{that it requires} substantial amounts of parallel speech-and-motion training data of sufficient quality in order to obtain good prediction performance.
In the future, we might overcome this limitation by obtaining datasets directly from publicly-available video recordings using motion-estimation techniques.


\subsection{Future work}
\label{ssec:future_work}
\addcontentsline{toc}{subsection}{Introduction}
We see several interesting directions for future \hk{research}:

\hk{Firstly, it is beneficial to} make the model probabilistic, e.g., by using a \hk{Variational Autoencoder (VAE) as in \cite{kucherenko2018data}.
A person is likely to gesticulate differently at different times for the same utterance}. It is thus an appealing idea to make a conversational agent also generate different gestures every time they speak the same sentence. \rmv{For this we need to make the mapping probabilistic, to represent a probability distribution over plausible motions and then draw samples from that distribution. VAEs can provide us with this functionality.}
    
\hk{Secondly, text should be taken} into account, e.g., as in \cite{ishii2018generating}.
Gestures that co-occur with speech depend \hk{greatly} on the semantic content of the utterance. Our model generates mostly beat gestures, as we rely only on speech acoustics as input. Hence \hk{the model} can benefit from incorporating the text transcription of the utterance \hk{along with the speech audio}. This may enable producing a wider range of gestures (also metaphoric and deictic gestures).
    
Lastly, the learned model can be applied to a humanoid robot so that the robot's speech is accompanied by appropriate co-speech gestures, for instance on the NAO robot as in  \cite{yoon2018robots}.

\section*{Acknowledgement}
The authors would like to thank Sanne van Waveren, Iolanda Leite and Simon Alexanderson for helpful discussions. 
This project is supported by the Swedish Foundation for Strategic Research Grant No.: RIT15-0107 (EACare). This work was also partially supported by JSPS Grant-in-Aid for Young Scientists (B) Grant Number 17K18075.

\break
\balance

%

%
\bibliographystyle{ACM-Reference-Format}
\bibliography{sample-bibliography}


\begin{thebibliography}{00}


\ifx \showCODEN    \undefined \def \showCODEN     #1{\unskip}     \fi
\ifx \showDOI      \undefined \def \showDOI       #1{#1}\fi
\ifx \showISBNx    \undefined \def \showISBNx     #1{\unskip}     \fi
\ifx \showISBNxiii \undefined \def \showISBNxiii  #1{\unskip}     \fi
\ifx \showISSN     \undefined \def \showISSN      #1{\unskip}     \fi
\ifx \showLCCN     \undefined \def \showLCCN      #1{\unskip}     \fi
\ifx \shownote     \undefined \def \shownote      #1{#1}          \fi
\ifx \showarticletitle \undefined \def \showarticletitle #1{#1}   \fi
\ifx \showURL      \undefined \def \showURL       {\relax}        \fi
\providecommand\bibfield[2]{#2}
\providecommand\bibinfo[2]{#2}
\providecommand\natexlab[1]{#1}
\providecommand\showeprint[2][]{arXiv:#2}

\bibitem[\protect\citeauthoryear{Bergmann and Kopp}{Bergmann and Kopp}{2009}]%
        {bergmann2009GNetIc}
\bibfield{author}{\bibinfo{person}{Kirsten Bergmann} {and}
  \bibinfo{person}{Stefan Kopp}.} \bibinfo{year}{2009}\natexlab{}.
\newblock \showarticletitle{{GNetIc}--Using {B}ayesian decision networks for
  iconic gesture generation}. In \bibinfo{booktitle}{{\em International
  Workshop on Intelligent Virtual Agents (IVA '09)}}. Springer,
  \bibinfo{pages}{76--89}.
\newblock


\bibitem[\protect\citeauthoryear{Boersma}{Boersma}{2002}]%
        {boersma2002praat}
\bibfield{author}{\bibinfo{person}{Paul Boersma}.}
  \bibinfo{year}{2002}\natexlab{}.
\newblock \showarticletitle{Praat, a system for doing phonetics by computer}.
\newblock \bibinfo{journal}{{\em Glot International\/}} \bibinfo{volume}{5},
  \bibinfo{number}{9/10} (\bibinfo{year}{2002}), \bibinfo{pages}{341--345}.
\newblock


\bibitem[\protect\citeauthoryear{Breazeal, Kidd, Thomaz, Hoffman, and
  Berlin}{Breazeal et~al\mbox{.}}{2005}]%
        {breazeal2005effects}
\bibfield{author}{\bibinfo{person}{Cynthia Breazeal}, \bibinfo{person}{Cory~D
  Kidd}, \bibinfo{person}{Andrea~Lockerd Thomaz}, \bibinfo{person}{Guy
  Hoffman}, {and} \bibinfo{person}{Matt Berlin}.}
  \bibinfo{year}{2005}\natexlab{}.
\newblock \showarticletitle{Effects of nonverbal communication on efficiency
  and robustness in human-robot teamwork}. In \bibinfo{booktitle}{{\em
  International Conference on Intelligent Robots and Systems, (IROS '05)}}.
  IEEE, \bibinfo{pages}{708--713}.
\newblock


\bibitem[\protect\citeauthoryear{B{\"u}tepage, Black, Kragic, and
  Kjellstr{\"o}m}{B{\"u}tepage et~al\mbox{.}}{2017}]%
        {butepage2017deep}
\bibfield{author}{\bibinfo{person}{Judith B{\"u}tepage},
  \bibinfo{person}{Michael~J. Black}, \bibinfo{person}{Danica Kragic}, {and}
  \bibinfo{person}{Hedvig Kjellstr{\"o}m}.} \bibinfo{year}{2017}\natexlab{}.
\newblock \showarticletitle{Deep representation learning for human notion
  prediction and classification}. In \bibinfo{booktitle}{{\em IEEE Conference
  on Computer Vision and Pattern Recognition (CVPR '17)}}. IEEE.
\newblock


\bibitem[\protect\citeauthoryear{Cassell, Vilhj{\'a}lmsson, and
  Bickmore}{Cassell et~al\mbox{.}}{2001}]%
        {cassell2001beat}
\bibfield{author}{\bibinfo{person}{Justine Cassell},
  \bibinfo{person}{Hannes~H{\"o}gni Vilhj{\'a}lmsson}, {and}
  \bibinfo{person}{Timothy Bickmore}.} \bibinfo{year}{2001}\natexlab{}.
\newblock \showarticletitle{Beat: The behavior expression animation toolkit}.
  In \bibinfo{booktitle}{{\em Annual Conference on Computer Graphics and
  Interactive Techniques (SIGGRAPH '01)}}. ACM.
\newblock


\bibitem[\protect\citeauthoryear{Chiu and Marsella}{Chiu and Marsella}{2011}]%
        {chiu2011train}
\bibfield{author}{\bibinfo{person}{Chung-Cheng Chiu} {and}
  \bibinfo{person}{Stacy Marsella}.} \bibinfo{year}{2011}\natexlab{}.
\newblock \showarticletitle{How to train your avatar: {A} data driven approach
  to gesture generation}. In \bibinfo{booktitle}{{\em International Workshop on
  Intelligent Virtual Agents (IVA'11)}}. Springer, \bibinfo{pages}{127--140}.
\newblock


\bibitem[\protect\citeauthoryear{Chiu, Morency, and Marsella}{Chiu
  et~al\mbox{.}}{2015}]%
        {chiu2015predicting}
\bibfield{author}{\bibinfo{person}{Chung-Cheng Chiu},
  \bibinfo{person}{Louis-Philippe Morency}, {and} \bibinfo{person}{Stacy
  Marsella}.} \bibinfo{year}{2015}\natexlab{}.
\newblock \showarticletitle{Predicting co-verbal gestures: A deep and temporal
  modeling approach}. In \bibinfo{booktitle}{{\em International Conference on
  Intelligent Virtual Agents (IVA '15)}}. Springer.
\newblock


\bibitem[\protect\citeauthoryear{Cho, van Merri{\"e}nboer, Bahdanau, and
  Bengio}{Cho et~al\mbox{.}}{2014}]%
        {cho2014properties}
\bibfield{author}{\bibinfo{person}{Kyunghyun Cho}, \bibinfo{person}{Bart van
  Merri{\"e}nboer}, \bibinfo{person}{Dzmitry Bahdanau}, {and}
  \bibinfo{person}{Yoshua Bengio}.} \bibinfo{year}{2014}\natexlab{}.
\newblock \showarticletitle{On the properties of neural machine translation:
  Encoder--decoder approaches}.
\newblock \bibinfo{journal}{{\em Syntax, Semantics and Structure in Statistical
  Translation\/}} (\bibinfo{year}{2014}), \bibinfo{pages}{103}.
\newblock


\bibitem[\protect\citeauthoryear{Davis and Mermelstein}{Davis and
  Mermelstein}{1980}]%
        {davis1980comparison}
\bibfield{author}{\bibinfo{person}{Steven Davis} {and} \bibinfo{person}{Paul
  Mermelstein}.} \bibinfo{year}{1980}\natexlab{}.
\newblock \showarticletitle{Comparison of parametric representations for
  monosyllabic word recognition in continuously spoken sentences}.
\newblock \bibinfo{journal}{{\em IEEE International Conference on Acoustics,
  Speech and Signal Processing (ACASSP '80)\/}} \bibinfo{volume}{28},
  \bibinfo{number}{4}, \bibinfo{pages}{357--366}.
\newblock


\bibitem[\protect\citeauthoryear{Greenwood, Laycock, and Matthews}{Greenwood
  et~al\mbox{.}}{2017}]%
        {greenwood2017predicting}
\bibfield{author}{\bibinfo{person}{David Greenwood}, \bibinfo{person}{Stephen
  Laycock}, {and} \bibinfo{person}{Iain Matthews}.}
  \bibinfo{year}{2017}\natexlab{}.
\newblock \showarticletitle{Predicting head pose from speech with a conditional
  variational autoencoder}. In \bibinfo{booktitle}{{\em Conference of the
  International Speech Communication Association (Interspeech '17)}}. ISCA,
  \bibinfo{pages}{3991--3995}.
\newblock


\bibitem[\protect\citeauthoryear{Haag and Shimodaira}{Haag and
  Shimodaira}{2016}]%
        {haag2016bidirectional}
\bibfield{author}{\bibinfo{person}{Kathrin Haag} {and} \bibinfo{person}{Hiroshi
  Shimodaira}.} \bibinfo{year}{2016}\natexlab{}.
\newblock \showarticletitle{Bidirectional {LSTM} networks employing stacked
  bottleneck features for expressive speech-driven head motion synthesis}. In
  \bibinfo{booktitle}{{\em International Conference on Intelligent Virtual
  Agents (IVA '16)}}. Springer, \bibinfo{pages}{198--207}.
\newblock


\bibitem[\protect\citeauthoryear{Habibie, Holden, Schwarz, Yearsley, and
  Komura}{Habibie et~al\mbox{.}}{2017}]%
        {habibie2017recurrent}
\bibfield{author}{\bibinfo{person}{Ikhsanul Habibie}, \bibinfo{person}{Daniel
  Holden}, \bibinfo{person}{Jonathan Schwarz}, \bibinfo{person}{Joe Yearsley},
  {and} \bibinfo{person}{Taku Komura}.} \bibinfo{year}{2017}\natexlab{}.
\newblock \showarticletitle{A recurrent variational autoencoder for human
  motion synthesis}.
\newblock \bibinfo{journal}{{\em IEEE Computer Graphics and Applications\/}}
  \bibinfo{volume}{37} (\bibinfo{year}{2017}), \bibinfo{pages}{4}.
\newblock


\bibitem[\protect\citeauthoryear{Hasegawa, Kaneko, Shirakawa, Sakuta, and
  Sumi}{Hasegawa et~al\mbox{.}}{2018}]%
        {hasegawa2018evaluation}
\bibfield{author}{\bibinfo{person}{Dai Hasegawa}, \bibinfo{person}{Naoshi
  Kaneko}, \bibinfo{person}{Shinichi Shirakawa}, \bibinfo{person}{Hiroshi
  Sakuta}, {and} \bibinfo{person}{Kazuhiko Sumi}.}
  \bibinfo{year}{2018}\natexlab{}.
\newblock \showarticletitle{Evaluation of speech-to-gesture generation using
  bi-directional LSTM network}. In \bibinfo{booktitle}{{\em International
  Conference on Intelligent Virtual Agents (IVA '18)}}. ACM,
  \bibinfo{pages}{79--86}.
\newblock


\bibitem[\protect\citeauthoryear{Holden, Saito, and Komura}{Holden
  et~al\mbox{.}}{2016}]%
        {holden2016deep}
\bibfield{author}{\bibinfo{person}{Daniel Holden}, \bibinfo{person}{Jun Saito},
  {and} \bibinfo{person}{Taku Komura}.} \bibinfo{year}{2016}\natexlab{}.
\newblock \showarticletitle{A deep learning framework for character motion
  synthesis and editing}.
\newblock \bibinfo{journal}{{\em ACM Transactions on Graphics (TOG)\/}}
  \bibinfo{volume}{35}, \bibinfo{number}{4} (\bibinfo{year}{2016}),
  \bibinfo{pages}{138:1--138:11}.
\newblock


\bibitem[\protect\citeauthoryear{Holden, Saito, Komura, and Joyce}{Holden
  et~al\mbox{.}}{2015}]%
        {holden2015learning}
\bibfield{author}{\bibinfo{person}{Daniel Holden}, \bibinfo{person}{Jun Saito},
  \bibinfo{person}{Taku Komura}, {and} \bibinfo{person}{Thomas Joyce}.}
  \bibinfo{year}{2015}\natexlab{}.
\newblock \showarticletitle{Learning motion manifolds with convolutional
  autoencoders}. In \bibinfo{booktitle}{{\em SIGGRAPH Asia Technical Briefs}}.
  \bibinfo{pages}{18:1--18:4}.
\newblock


\bibitem[\protect\citeauthoryear{Huang and Mutlu}{Huang and Mutlu}{2012}]%
        {huang2012robot}
\bibfield{author}{\bibinfo{person}{Chien-Ming Huang} {and}
  \bibinfo{person}{Bilge Mutlu}.} \bibinfo{year}{2012}\natexlab{}.
\newblock \showarticletitle{Robot behavior toolkit: Generating effective social
  behaviors for robots}. In \bibinfo{booktitle}{{\em International Conference
  on Human Robot Interaction (HRI '12)}}. ACM/IEEE.
\newblock


\bibitem[\protect\citeauthoryear{Ishii, Katayama, Higashinaka, and
  Tomita}{Ishii et~al\mbox{.}}{2018}]%
        {ishii2018generating}
\bibfield{author}{\bibinfo{person}{Ryo Ishii}, \bibinfo{person}{Taichi
  Katayama}, \bibinfo{person}{Ryuichiro Higashinaka}, {and}
  \bibinfo{person}{Junji Tomita}.} \bibinfo{year}{2018}\natexlab{}.
\newblock \showarticletitle{Generating body motions using spoken language in
  dialogue}. In \bibinfo{booktitle}{{\em Proceedings of the 18th International
  Conference on Intelligent Virtual Agents}} {\em (\bibinfo{series}{IVA '18})}.
  \bibinfo{publisher}{ACM}, \bibinfo{pages}{87--92}.
\newblock
\showISBNx{978-1-4503-6013-5}


\bibitem[\protect\citeauthoryear{Kingma and Ba}{Kingma and Ba}{2015}]%
        {kingma2014adam}
\bibfield{author}{\bibinfo{person}{Diederik~P Kingma} {and}
  \bibinfo{person}{Jimmy Ba}.} \bibinfo{year}{2015}\natexlab{}.
\newblock \showarticletitle{Adam: A method for stochastic optimization}. In
  \bibinfo{booktitle}{{\em International Conference on Learning
  Representations}} {\em (\bibinfo{series}{ICLR '15})}.
\newblock


\bibitem[\protect\citeauthoryear{Knapp, Hall, and Horgan}{Knapp
  et~al\mbox{.}}{2013}]%
        {knapp2013nonverbal}
\bibfield{author}{\bibinfo{person}{Mark~L Knapp}, \bibinfo{person}{Judith~A
  Hall}, {and} \bibinfo{person}{Terrence~G Horgan}.}
  \bibinfo{year}{2013}\natexlab{}.
\newblock \bibinfo{booktitle}{{\em Nonverbal Communication in Human
  Interaction}}.
\newblock \bibinfo{publisher}{Wadsworth, Cengage Learning}.
\newblock


\bibitem[\protect\citeauthoryear{Kucherenko}{Kucherenko}{2018}]%
        {kucherenko2018data}
\bibfield{author}{\bibinfo{person}{Taras Kucherenko}.}
  \bibinfo{year}{2018}\natexlab{}.
\newblock \showarticletitle{Data driven non-verbal behavior generation for
  humanoid robots}. In \bibinfo{booktitle}{{\em ACM International Conference on
  Multimodal Interaction, Doctoral Consortium}} {\em (\bibinfo{series}{ICMI
  '18})}. \bibinfo{publisher}{ACM}, \bibinfo{pages}{520--523}.
\newblock
\showISBNx{978-1-4503-5692-3}


\bibitem[\protect\citeauthoryear{Kucherenko, Hasegawa, Kaneko, Henter, and
  Kjellstr{\"o}m}{Kucherenko et~al\mbox{.}}{2019}]%
        {kucherenko2019importance}
\bibfield{author}{\bibinfo{person}{Taras Kucherenko}, \bibinfo{person}{Dai
  Hasegawa}, \bibinfo{person}{Naoshi Kaneko}, \bibinfo{person}{Gustav~Eje
  Henter}, {and} \bibinfo{person}{Hedvig Kjellstr{\"o}m}.}
  \bibinfo{year}{2019}\natexlab{}.
\newblock \showarticletitle{On the importance of representations for
  speech-driven gesture generation}. In \bibinfo{booktitle}{{\em International
  Conference on Autonomous Agents and Multiagent Systems}} {\em
  (\bibinfo{series}{AAMAS '19})}. \bibinfo{publisher}{ACM}.
\newblock


\bibitem[\protect\citeauthoryear{Levine, Kr{\"a}henb{\"u}hl, Thrun, and
  Koltun}{Levine et~al\mbox{.}}{2010}]%
        {levine2010gesture}
\bibfield{author}{\bibinfo{person}{Sergey Levine}, \bibinfo{person}{Philipp
  Kr{\"a}henb{\"u}hl}, \bibinfo{person}{Sebastian Thrun}, {and}
  \bibinfo{person}{Vladlen Koltun}.} \bibinfo{year}{2010}\natexlab{}.
\newblock \showarticletitle{Gesture controllers}.
\newblock \bibinfo{journal}{{\em ACM Transactions on Graphics (TOG)\/}}
  \bibinfo{volume}{29}, \bibinfo{number}{4} (\bibinfo{year}{2010}),
  \bibinfo{pages}{124}.
\newblock


\bibitem[\protect\citeauthoryear{Liu and Taniguchi}{Liu and Taniguchi}{2014}]%
        {liu2014feature}
\bibfield{author}{\bibinfo{person}{Hailong Liu} {and} \bibinfo{person}{Tadahiro
  Taniguchi}.} \bibinfo{year}{2014}\natexlab{}.
\newblock \showarticletitle{Feature extraction and pattern recognition for
  human motion by a deep sparse autoencoder}. In \bibinfo{booktitle}{{\em
  International Conference on Computer and Information Technology (CIT '14)}}.
  IEEE, \bibinfo{pages}{173--181}.
\newblock


\bibitem[\protect\citeauthoryear{Martinez, Black, and Romero}{Martinez
  et~al\mbox{.}}{2017}]%
        {martinez2017human}
\bibfield{author}{\bibinfo{person}{Julieta Martinez},
  \bibinfo{person}{Michael~J Black}, {and} \bibinfo{person}{Javier Romero}.}
  \bibinfo{year}{2017}\natexlab{}.
\newblock \showarticletitle{On human motion prediction using recurrent neural
  networks}. In \bibinfo{booktitle}{{\em IEEE Conference on Computer Vision and
  Pattern Recognition (CVPR '17)}}. IEEE, \bibinfo{pages}{4674--4683}.
\newblock


\bibitem[\protect\citeauthoryear{Matsumoto, Frank, and Hwang}{Matsumoto
  et~al\mbox{.}}{2013}]%
        {matsumoto2013nonverbal}
\bibfield{author}{\bibinfo{person}{David Matsumoto}, \bibinfo{person}{Mark~G
  Frank}, {and} \bibinfo{person}{Hyi~Sung Hwang}.}
  \bibinfo{year}{2013}\natexlab{}.
\newblock \bibinfo{booktitle}{{\em Nonverbal Communication: Science and
  Applications}}.
\newblock \bibinfo{publisher}{Sage}.
\newblock


\bibitem[\protect\citeauthoryear{McNeill}{McNeill}{1992}]%
        {mcneill1992hand}
\bibfield{author}{\bibinfo{person}{David McNeill}.}
  \bibinfo{year}{1992}\natexlab{}.
\newblock \bibinfo{booktitle}{{\em Hand and Mind: What Gestures Reveal about
  Thought}}.
\newblock \bibinfo{publisher}{University of Chicago Press}.
\newblock


\bibitem[\protect\citeauthoryear{Ng-Thow-Hing, Luo, and Okita}{Ng-Thow-Hing
  et~al\mbox{.}}{2010}]%
        {ng2010synchronized}
\bibfield{author}{\bibinfo{person}{Victor Ng-Thow-Hing},
  \bibinfo{person}{Pengcheng Luo}, {and} \bibinfo{person}{Sandra Okita}.}
  \bibinfo{year}{2010}\natexlab{}.
\newblock \showarticletitle{Synchronized gesture and speech production for
  humanoid robots}. In \bibinfo{booktitle}{{\em International Conference on
  Intelligent Robots and Systems}} {\em (\bibinfo{series}{IROS '10})}.
  \bibinfo{publisher}{IEEE/RSJ}.
\newblock


\bibitem[\protect\citeauthoryear{Pavllo, Grangier, and Auli}{Pavllo
  et~al\mbox{.}}{2018}]%
        {pavllo2018quaternet}
\bibfield{author}{\bibinfo{person}{Dario Pavllo}, \bibinfo{person}{David
  Grangier}, {and} \bibinfo{person}{Michael Auli}.}
  \bibinfo{year}{2018}\natexlab{}.
\newblock \showarticletitle{{Q}uater{N}et: {A} quaternion-based recurrent model
  for human motion}. In \bibinfo{booktitle}{{\em British Machine Vision
  Conference}} {\em (\bibinfo{series}{BMVC '18})}.
\newblock


\bibitem[\protect\citeauthoryear{Sadoughi and Busso}{Sadoughi and
  Busso}{2017}]%
        {sadoughi2017joint}
\bibfield{author}{\bibinfo{person}{Najmeh Sadoughi} {and}
  \bibinfo{person}{Carlos Busso}.} \bibinfo{year}{2017}\natexlab{}.
\newblock \showarticletitle{Joint learning of speech-driven facial motion with
  bidirectional long-short term memory}. In \bibinfo{booktitle}{{\em
  International Conference on Intelligent Virtual Agents}} {\em
  (\bibinfo{series}{IVA '17})}. Springer, \bibinfo{pages}{389--402}.
\newblock


\bibitem[\protect\citeauthoryear{Sadoughi and Busso}{Sadoughi and
  Busso}{2018}]%
        {sadoughi2018novel}
\bibfield{author}{\bibinfo{person}{Najmeh Sadoughi} {and}
  \bibinfo{person}{Carlos Busso}.} \bibinfo{year}{2018}\natexlab{}.
\newblock \showarticletitle{Novel realizations of speech-driven head movements
  with generative adversarial networks}. In \bibinfo{booktitle}{{\em IEEE
  International Conference on Acoustics, Speech and Signal Processing (ICASSP
  '18)}}. IEEE, \bibinfo{pages}{6169--6173}.
\newblock


\bibitem[\protect\citeauthoryear{Sadoughi and Busso}{Sadoughi and
  Busso}{2019}]%
        {sadoughi2017speech}
\bibfield{author}{\bibinfo{person}{Najmeh Sadoughi} {and}
  \bibinfo{person}{Carlos Busso}.} \bibinfo{year}{2019}\natexlab{}.
\newblock \showarticletitle{Speech-driven animation with meaningful behaviors}.
\newblock \bibinfo{journal}{{\em Speech Communication\/}}
  \bibinfo{volume}{110} (\bibinfo{year}{2019}), \bibinfo{pages}{90--100}.
\newblock


\bibitem[\protect\citeauthoryear{Salem, Eyssel, Rohlfing, Kopp, and
  Joublin}{Salem et~al\mbox{.}}{2013}]%
        {salem2013err}
\bibfield{author}{\bibinfo{person}{Maha Salem}, \bibinfo{person}{Friederike
  Eyssel}, \bibinfo{person}{Katharina Rohlfing}, \bibinfo{person}{Stefan Kopp},
  {and} \bibinfo{person}{Frank Joublin}.} \bibinfo{year}{2013}\natexlab{}.
\newblock \showarticletitle{To err is human (-like): Effects of robot gesture
  on perceived anthropomorphism and likability}.
\newblock \bibinfo{journal}{{\em International Journal of Social Robotics\/}}
  \bibinfo{volume}{5}, \bibinfo{number}{3} (\bibinfo{year}{2013}),
  \bibinfo{pages}{313--323}.
\newblock


\bibitem[\protect\citeauthoryear{Salvi, Beskow, Al~Moubayed, and
  Granstr{\"o}m}{Salvi et~al\mbox{.}}{2009}]%
        {salvi2009synface}
\bibfield{author}{\bibinfo{person}{Giampiero Salvi}, \bibinfo{person}{Jonas
  Beskow}, \bibinfo{person}{Samer Al~Moubayed}, {and}
  \bibinfo{person}{Bj{\"o}rn Granstr{\"o}m}.} \bibinfo{year}{2009}\natexlab{}.
\newblock \showarticletitle{SynFace: Speech-driven facial animation for virtual
  speech-reading support}.
\newblock \bibinfo{journal}{{\em EURASIP Journal on Audio, Speech, and Music
  Processing\/}} (\bibinfo{year}{2009}), \bibinfo{pages}{3}.
\newblock


\bibitem[\protect\citeauthoryear{Suwajanakorn, Seitz, and
  Kemelmacher-Shlizerman}{Suwajanakorn et~al\mbox{.}}{2017}]%
        {suwajanakorn2017synthesizing}
\bibfield{author}{\bibinfo{person}{Supasorn Suwajanakorn},
  \bibinfo{person}{Steven~M Seitz}, {and} \bibinfo{person}{Ira
  Kemelmacher-Shlizerman}.} \bibinfo{year}{2017}\natexlab{}.
\newblock \showarticletitle{Synthesizing Obama: Learning lip sync from audio}.
\newblock \bibinfo{journal}{{\em ACM Transactions on Graphics (TOG)\/}}
  \bibinfo{volume}{36}, \bibinfo{number}{4} (\bibinfo{year}{2017}),
  \bibinfo{pages}{95}.
\newblock


\bibitem[\protect\citeauthoryear{Takeuchi, Hasegawa, Shirakawa, Kaneko, Sakuta,
  and Sumi}{Takeuchi et~al\mbox{.}}{2017a}]%
        {takeuchi2017speech}
\bibfield{author}{\bibinfo{person}{Kenta Takeuchi}, \bibinfo{person}{Dai
  Hasegawa}, \bibinfo{person}{Shinichi Shirakawa}, \bibinfo{person}{Naoshi
  Kaneko}, \bibinfo{person}{Hiroshi Sakuta}, {and} \bibinfo{person}{Kazuhiko
  Sumi}.} \bibinfo{year}{2017}\natexlab{a}.
\newblock \showarticletitle{Speech-to-gesture generation: A challenge in deep
  learning approach with bi-directional LSTM}. In \bibinfo{booktitle}{{\em
  International Conference on Human Agent Interaction (HAI '17)}}.
\newblock


\bibitem[\protect\citeauthoryear{Takeuchi, Kubota, Suzuki, Hasegawa, and
  Sakuta}{Takeuchi et~al\mbox{.}}{2017b}]%
        {takeuchi2017creating}
\bibfield{author}{\bibinfo{person}{Kenta Takeuchi}, \bibinfo{person}{Souichirou
  Kubota}, \bibinfo{person}{Keisuke Suzuki}, \bibinfo{person}{Dai Hasegawa},
  {and} \bibinfo{person}{Hiroshi Sakuta}.} \bibinfo{year}{2017}\natexlab{b}.
\newblock \showarticletitle{Creating a gesture-speech dataset for speech-based
  automatic gesture generation}. In \bibinfo{booktitle}{{\em International
  Conference on Human-Computer Interaction}} {\em (\bibinfo{series}{HCI '17})}.
  Springer, \bibinfo{pages}{198--202}.
\newblock


\bibitem[\protect\citeauthoryear{Vincent, Larochelle, Lajoie, Bengio, and
  Manzagol}{Vincent et~al\mbox{.}}{2010}]%
        {vincent2010stacked}
\bibfield{author}{\bibinfo{person}{Pascal Vincent}, \bibinfo{person}{Hugo
  Larochelle}, \bibinfo{person}{Isabelle Lajoie}, \bibinfo{person}{Yoshua
  Bengio}, {and} \bibinfo{person}{Pierre-Antoine Manzagol}.}
  \bibinfo{year}{2010}\natexlab{}.
\newblock \showarticletitle{Stacked denoising autoencoders: Learning useful
  representations in a deep network with a local denoising criterion}.
\newblock \bibinfo{journal}{{\em Journal of Machine Learning Research\/}}
  \bibinfo{volume}{11}, \bibinfo{number}{Dec} (\bibinfo{year}{2010}),
  \bibinfo{pages}{3371--3408}.
\newblock


\bibitem[\protect\citeauthoryear{Wagner, Malisz, and Kopp}{Wagner
  et~al\mbox{.}}{2014}]%
        {wagner2014gesture}
\bibfield{author}{\bibinfo{person}{Petra Wagner}, \bibinfo{person}{Zofia
  Malisz}, {and} \bibinfo{person}{Stefan Kopp}.}
  \bibinfo{year}{2014}\natexlab{}.
\newblock \showarticletitle{Gesture and speech in interaction: An overview}.
\newblock \bibinfo{journal}{{\em Speech Communication\/}}  \bibinfo{volume}{57}
  (\bibinfo{year}{2014}), \bibinfo{pages}{209--232}.
\newblock


\bibitem[\protect\citeauthoryear{Yoon, Ko, Jang, Lee, Kim, and Lee}{Yoon
  et~al\mbox{.}}{2019}]%
        {yoon2018robots}
\bibfield{author}{\bibinfo{person}{Youngwoo Yoon}, \bibinfo{person}{Woo-Ri Ko},
  \bibinfo{person}{Minsu Jang}, \bibinfo{person}{Jaeyeon Lee},
  \bibinfo{person}{Jaehong Kim}, {and} \bibinfo{person}{Geehyuk Lee}.}
  \bibinfo{year}{2019}\natexlab{}.
\newblock \showarticletitle{Robots learn social skills: End-to-end learning of
  co-speech gesture generation for humanoid robots}. In
  \bibinfo{booktitle}{{\em International Conference on Robotics and Automation
  (ICRA '19)}}. IEEE.
\newblock


\bibitem[\protect\citeauthoryear{Zhang, Starke, Komura, and Saito}{Zhang
  et~al\mbox{.}}{2018}]%
        {zhang2018mode}
\bibfield{author}{\bibinfo{person}{He Zhang}, \bibinfo{person}{Sebastian
  Starke}, \bibinfo{person}{Taku Komura}, {and} \bibinfo{person}{Jun Saito}.}
  \bibinfo{year}{2018}\natexlab{}.
\newblock \showarticletitle{Mode-adaptive neural networks for quadruped motion
  control}.
\newblock \bibinfo{journal}{{\em ACM Transactions on Graphics (TOG)\/}}
  \bibinfo{volume}{37}, \bibinfo{number}{4} (\bibinfo{year}{2018}),
  \bibinfo{pages}{145}.
\newblock


\bibitem[\protect\citeauthoryear{Zhou, Li, Xiao, He, Huang, and Li}{Zhou
  et~al\mbox{.}}{2017}]%
        {li2017auto}
\bibfield{author}{\bibinfo{person}{Yi Zhou}, \bibinfo{person}{Zimo Li},
  \bibinfo{person}{Shuangjiu Xiao}, \bibinfo{person}{Chong He},
  \bibinfo{person}{Zeng Huang}, {and} \bibinfo{person}{Hao Li}.}
  \bibinfo{year}{2017}\natexlab{}.
\newblock \showarticletitle{Auto-conditioned recurrent networks for extended
  complex human motion synthesis}. In \bibinfo{booktitle}{{\em International
  Conference on Learning Representations}} {\em (\bibinfo{series}{ICLR '17})}.
\newblock


\end{thebibliography}

\end{document}